\documentclass[%
 reprint,
 amsmath,amssymb,
 showpacs,
 prd
]{revtex4-1}

\usepackage{graphicx}
\usepackage{dcolumn}
\usepackage{bm}
\usepackage{verbatim}
\usepackage{array}
\usepackage{amsmath}
\usepackage{lineno}
\usepackage[bookmarksnumbered, pdfstartview=FitH,colorlinks,urlcolor=blue, citecolor=blue,linkcolor=blue,]{hyperref}
\usepackage{setspace}
\usepackage{subfigure}
\usepackage{textcomp}

\newcommand{\jpsi}{J/\psi}
\newcommand{\psip}{\psi(3686)}

\begin{document}

\preprint{APS/123-QED}

\title{\boldmath Prospects to study hyperon-nucleon interactions at BESIII}

\author{Jianping Dai$^{1}$}
\author{Hai-Bo Li$^{2,3}$}
\email{lihb@ihep.ac.cn}
\author{Han Miao$^{2,3}$}
\email{miaohan@ihep.ac.cn}
\author{Jianyu Zhang$^{3}$}
\email{zhangjianyu@ucas.ac.cn}
\affiliation{%
        $^{1}$Department of Physics, Yunnan University, Kunming 650091, People's Republic of China\\
        $^{2}$Institute of High Energy Physics, Chinese Academy of Sciences, Beijing 100049, People's Republic of China\\
        $^{3}$University of Chinese Academy of Sciences, Beijing 100049, People's Republic of China
}%

\date{\today}

\begin{abstract}
The prospects to study hyperon-nuclei/nucleon interactions at BESIII and similar $e^+ e^-$ colliders are proposed in this work. Utilizing the large quantity of hyperons produced by the decay of 10 billion $J/\psi$ and 2.7 billion $\psi(3686)$ collected at BESIII, the cross sections of several specific elastic or inelastic hyperon-nuclei reactions can be measured via the scattering between the hyperons and the nucleus in the dense objects of BESIII detector. Subsequently, the cross sections of corresponding hyperon-nucleon interactions are able to be extracted with further phenomenological calculation. Furthermore, the interactions between antihyperon and nuclei/nucleon, including scattering and annihilation, can also be studied via the method proposed in this paper. The results will definitely benefit a lot the precise probe of the hyperon-nuclei/nucleon interactions and provide constraints for the studies of the potential of strong interaction, the origin of color confinement, the unified model for baryon-baryon interactions, and the internal structure of neutron stars. In addition, the desirable prospects of corresponding studies in the future Super Tau-Charm Factory (STCF) are discussed and estimated in this work.
\end{abstract}


\maketitle

\section{Introduction} \label{sec:intro}

Describing baryon-baryon interactions within a unified model has always been a challenge in both particle and nuclear physics~\cite{Vidana:2018bdi,Hiyama:2018lgs,Gal:2016boi,Tolos:2020aln}. Strong constraints and  well-established models exist for nucleon-nucleon interactions~\cite{Vidana:2018bdi,Hiyama:2018lgs}, but there are still difficulties in precisely modeling hyperon-nucleon scattering, especially hyperon-hyperon interactions, due to the lack of experimental measurements. Until now, there have only been a few measurements for hyperon-nucleon scattering~\cite{Eisele:1971mk,Sechi-Zorn:1968mao,Alexander:1968acu,Kadyk:1971tc,Hauptman:1977hr,KEK-PSE-251:1997cno,KEK-PS-E289:2000ytt,KEK-PSE289:2005nsj,Ahn:2005jz,J-PARCE40:2021qxa,J-PARCE40:2021bgw,CLAS:2021gur,J-PARCE40:2022nvq, BESIII:2023clq, BESIII:2023trh}, and only one for hyperon-hyperon scattering~\cite{ALICE:2022uso}, leaving theoretical models largely unconstrained~\cite{Haidenbauer:2005zh, Rijken:2010zzb, Polinder:2006zh,Haidenbauer:2013oca,Haidenbauer:2015zqb, Haidenbauer:2018gvg, Haidenbauer:2019boi, Haidenbauer:2023qhf, Li:2016paq, Li:2016mln, Ishii:2006ec,Ishii:2012ssm, Beane:2006gf,Beane:2010em, Schaefer:2005fi, Fujiwara:2006yh}.

The properties of hyperons in dense matter have attracted much interest due to their close connection with hypernuclei and the hyperon component in neutron stars~\cite{Tolos:2020aln}. Hyperons may exist within the inner layer of neutron stars whose structure strongly depends on the equation of state (EOS) of nuclear matter at supersaturation densities~\cite{Lattimer:2000nx}. The appearance of hyperons in the core softens the EOS, resulting in neutron stars with masses lower than 2$M_\odot$~\cite{Lonardoni:2014bwa}, where $M_\odot$ is the mass of the sun. However, studies based on observations from the LIGO and Virgo experiments~\cite{LIGOScientific:2018cki} indicate that the EOS can support neutron stars with masses above $1.97M_\odot$.  This is the so-called ``hyperon puzzle in neutron stars", warranting further experimental and theoretical studies on the hyperon-nucleon interaction.

Most of the previous measurements ($\Lambda p \to \Lambda p$, $\Sigma^- p\to \Sigma^- p, \Lambda n, \Sigma^0 n$, and $\Sigma^+ p\to \Sigma^+ p$) were accomplished in the bubble chamber era of the 1960s and 1970s~\cite{Eisele:1971mk,Sechi-Zorn:1968mao,Alexander:1968acu,Kadyk:1971tc,Hauptman:1977hr} using hyperons with momenta less than 1 GeV/c$^2$. After a gap of twenty years, an experimental group in KEK studied $\Sigma^+ p$ and $\Sigma^- p$ elastic scattering processes with a scintillating fiber block~\cite{KEK-PSE-251:1997cno,KEK-PS-E289:2000ytt, KEK-PSE289:2005nsj}, where the momentum of $\Sigma^{\pm}$ hyperon is within [0.35, 0.75] GeV/c$^2$, a little higher than that used in Ref~\cite{Eisele:1971mk}. Later, one group in KEK first measured the total cross section of $\Xi^- p\to \Lambda\Lambda$ reaction at $p_{\Xi} \sim 0.5$ GeV/c$^2$~\cite{Ahn:2005jz}. Then the E40 collaboration in J-PARC Hadron Experimental Facility updated the measurements of $\Sigma^{\pm} p\to \Sigma^{\pm} p$ and $\Sigma^- p \to \Lambda n$ scattering processes with 0.4 GeV/c$^2<p_{\Lambda}<0.85 $GeV/c$^2$~\cite{J-PARCE40:2021qxa,J-PARCE40:2021bgw,J-PARCE40:2022nvq}, and the CLAS collaboration performed an improved measurements on the cross section of $\Lambda p$ elastic scattering process with 0.9 GeV/c$^2<p_{\Lambda}<2.0 $GeV/c$^2$~\cite{CLAS:2021gur}. It should be mentioned that the uncertainties of all the above measurements are still large. Very recently, the reactions ${\Xi^0} + {^{9}{\rm Be}} \to {\Xi^-} + p + {^{8}{\rm Be}}$ and $\Lambda + {^{9}{\rm Be}} \to \Sigma^+ + X$ have been measured with $p_{\Xi^0} \approx 0.818~{\rm GeV}/c^{2}$ and $p_{\Lambda} \approx 1.074~{\rm GeV}/c^{2}$ by BESIII collaboration~\cite{BESIII:2023clq, BESIII:2023trh}, of which the method used is introduced in this paper.

On the theoretical side, many models have been proposed to describe the hyperon-nucleon and hyperon-hyperon interactions, including the meson-exchange model (with J\"{u}lich~\cite{Haidenbauer:2005zh} or Nijmegen~\cite{Rijken:2010zzb} potentials), chiral effective field theory ($\chi$EFT) approaches~\cite{Polinder:2006zh,Haidenbauer:2013oca,Haidenbauer:2015zqb, Haidenbauer:2018gvg, Haidenbauer:2019boi, Haidenbauer:2023qhf, Li:2016paq, Li:2016mln}, calculations on the lattice from HALQCD~\cite{Ishii:2006ec,Ishii:2012ssm} and NPLQCD~\cite{Beane:2006gf,Beane:2010em}, low-momentum models~\cite{Schaefer:2005fi} and quark model approaches~\cite{Fujiwara:2006yh}. The precision of the models above will definitely be further improved by more experimental measurements.

Experimental studies of hyperon-nucleon interactions still suffer a lot from the difficulty of obtaining a stable hyperon beam. Firstly, the lifetime of ground-state hyperons is usually of order $O(10^{-10})~{\rm s}$ due to the weak decay, which is too short to be a stable beam. Meanwhile, hyperons historically used for fixed-target experiments are commonly produced in the $K p$ and $\gamma p$ collisions, such as J-PARC and CLAS experiments, with a high hadronic background level. Compared with the fixed-target experiments, much more hyperons are accessible from the decay of charmonia produced at $e^+ e^-$ colliders, which have recently been used to measure the hyperon-nuclei interactions at BESIII~\cite{BESIII:2023clq, BESIII:2023trh}. Furthermore, abundant antihyperons produced in pair with hyperons bring exciting prospects to probe antihyperon-nuclei/nucleon interactions that have rarely been measured and studied.

In this work, the prospects of studying hyperon-nuclei/nucleon interactions at BESIII are discussed. The hyperon pairs from the decays of 10 billion $J/\psi$ and 2.7 billion $\psi(3686)$~\cite{BESIII:2020nme} provide an intense and high-quality hyperon beam, which makes it possible to study the hyperon-nuclei/nucleon interactions by the scattering between hyperons and the nucleus in the materials of the beam pipe and other supporting structures of the spectrometer.

Using this method, the cross sections of two hyperon-nuclei interactions have been measured by BESIII~\cite{BESIII:2023clq, BESIII:2023trh}. Subsequently, the cross sections of corresponding hyperon-nucleon reactions are able to be extracted with further phenomenological calculation. Furthermore, in the near future, the measurements will definitely benefit from the improved techniques and the larger datasets at the Super Tau-Charm Factory (STCF) in proposal~\cite{Achasov:2023gey}.

The following part will be a brief introduction of the expected measurements at BESIII and will take the recently measured~\cite{BESIII:2023trh} reaction $J/\psi \to \Lambda \bar{\Lambda}, \bar{\Lambda} \to \bar{p} \pi^+, \Lambda + {^{9}{\rm Be}} \to \Sigma^+ + X$ as an example.

\section{Expected measurement at BESIII}
\label{sec:BESIII}

\subsection{The BESIII detector and the target}

The BESIII detector is a magnetic spectrometer~\cite{BESIII:2009fln} located at the Beijing Electron Positron Collider (BEPCII). The cylindrical core of the BESIII detector consists of a helium-based  multilayer drift chamber (MDC), a plastic scintillator time-of-flight system (TOF), and a CsI(Tl) electromagnetic calorimeter (EMC), which are all enclosed in a superconducting solenoidal magnet providing a 1.0~T magnetic field.

In particularly, the BESIII detector has excellent performance on the reconstruction of long-lived particles, such as $K_S$ and ground-state hyperons ($\Lambda$, $\Sigma^{+,-}$, $\Xi^{0,-}$ and $\Omega^-$), and has published dozens of analyses involved with hyperon physics~\cite{BESIII:2022qax, BESIII:2021ypr, BESIII:2020fqg, BESIII:2018cnd, BESIII:2022lsz, BESIII:2023drj}. Up to now, BESIII has collected the largest data samples around the world within the designed energy region~\cite{BESIII:2020nme}.

\begin{figure}[htbp]
        \centering
        \includegraphics[width=0.45\textwidth]{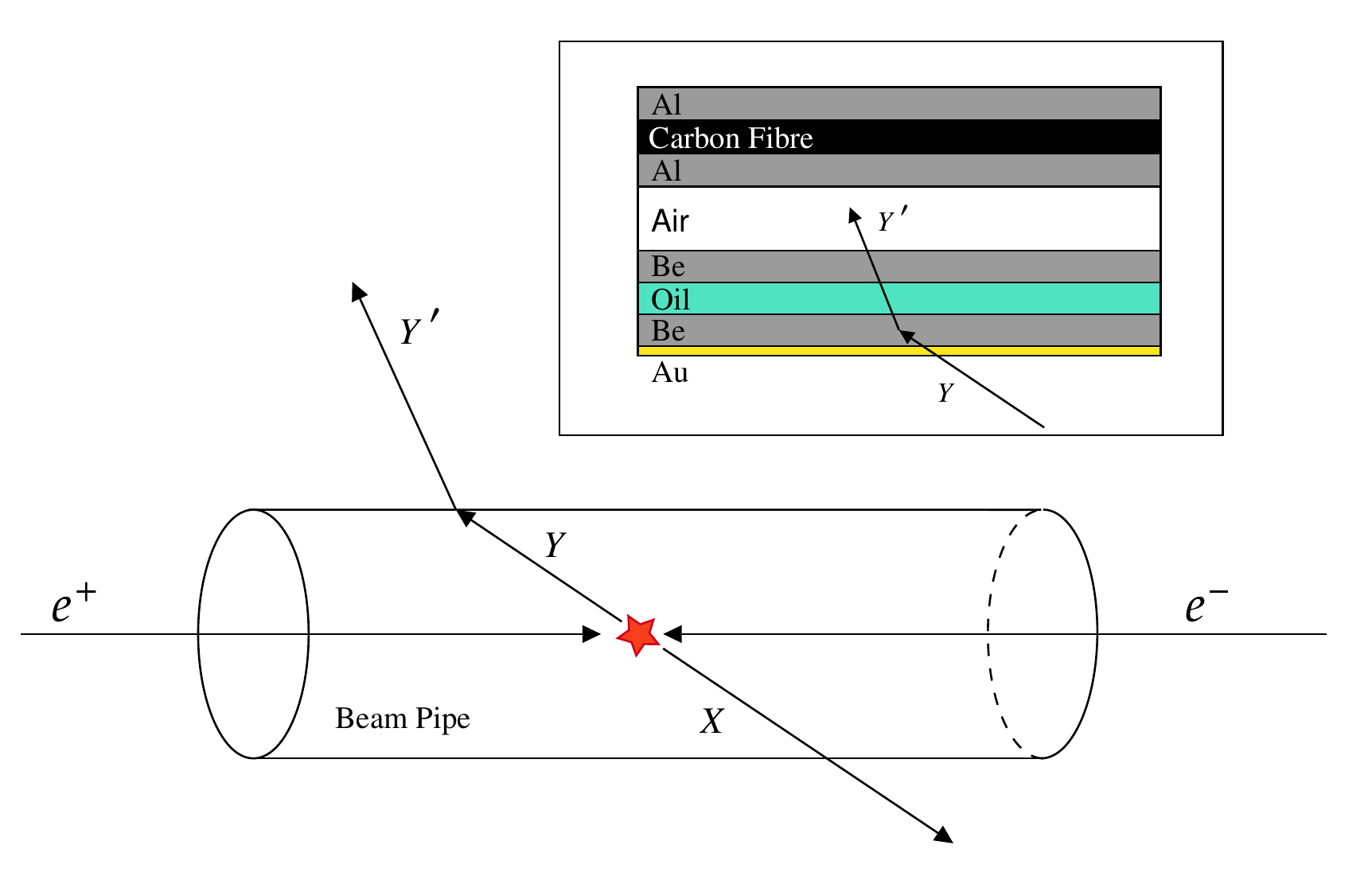}
	\caption{A schematic picture of hyperon-nucleon interactions at $e^+ e^-$ collider represented by BESIII. The symbol $Y$ denotes the hyperon interacting with the nucleus and $X$ represents the other particles produced in the $e^+ e^-$ collision together with $Y$. The symbol $Y^{\prime}$ denotes the particles produced in the hyperon-nucleus interaction. The structure of the target is shown in the top right corner. See Fig.~\ref{fig:interaction} for a more detailed view of the target.}
        \label{fig:r99bes}
\end{figure}

\begin{figure}[htbp]
        \centering
        \subfigure[]{ \includegraphics[width=0.35\textwidth]{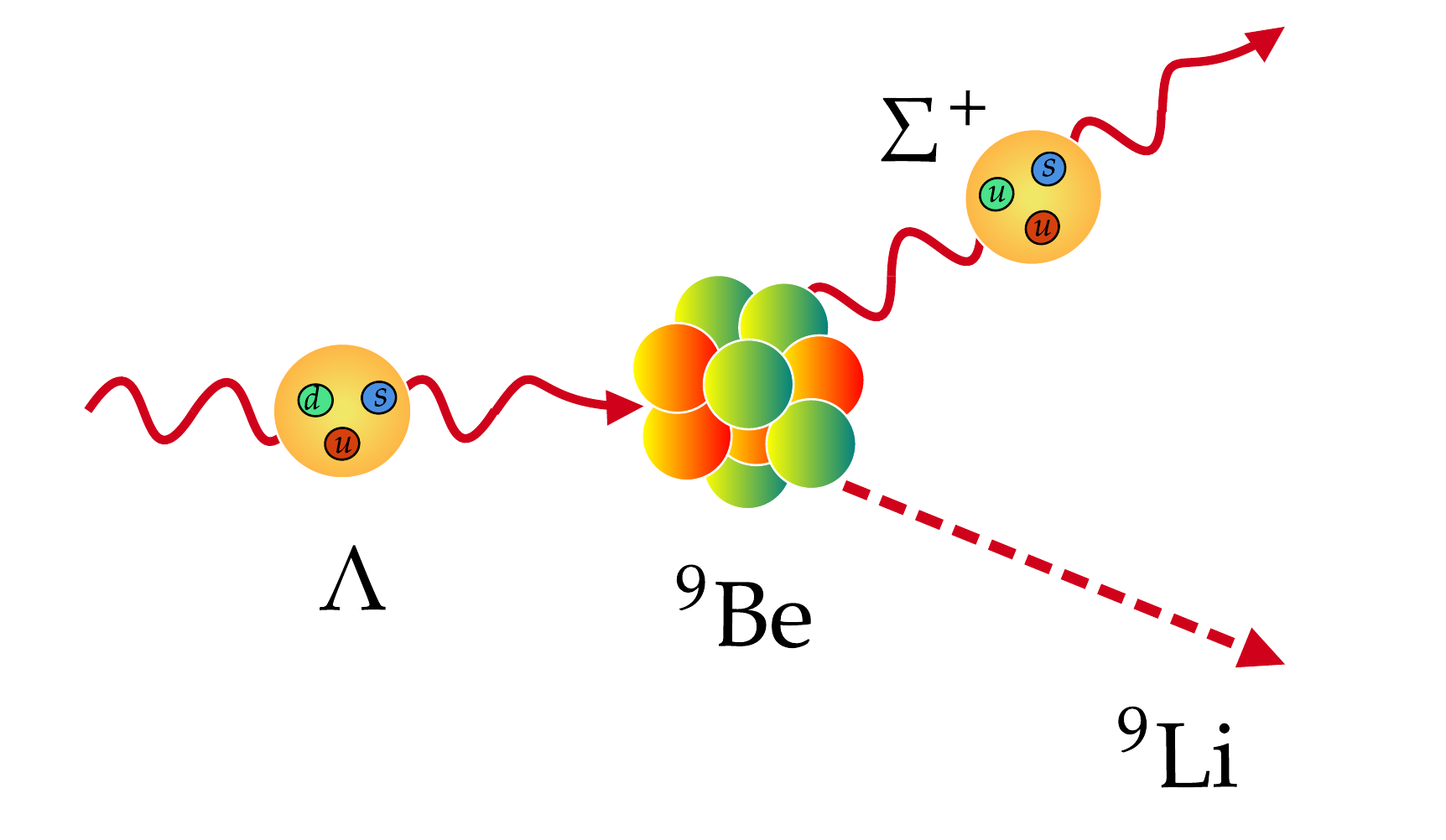} }
        \subfigure[]{ \includegraphics[width=0.45\textwidth]{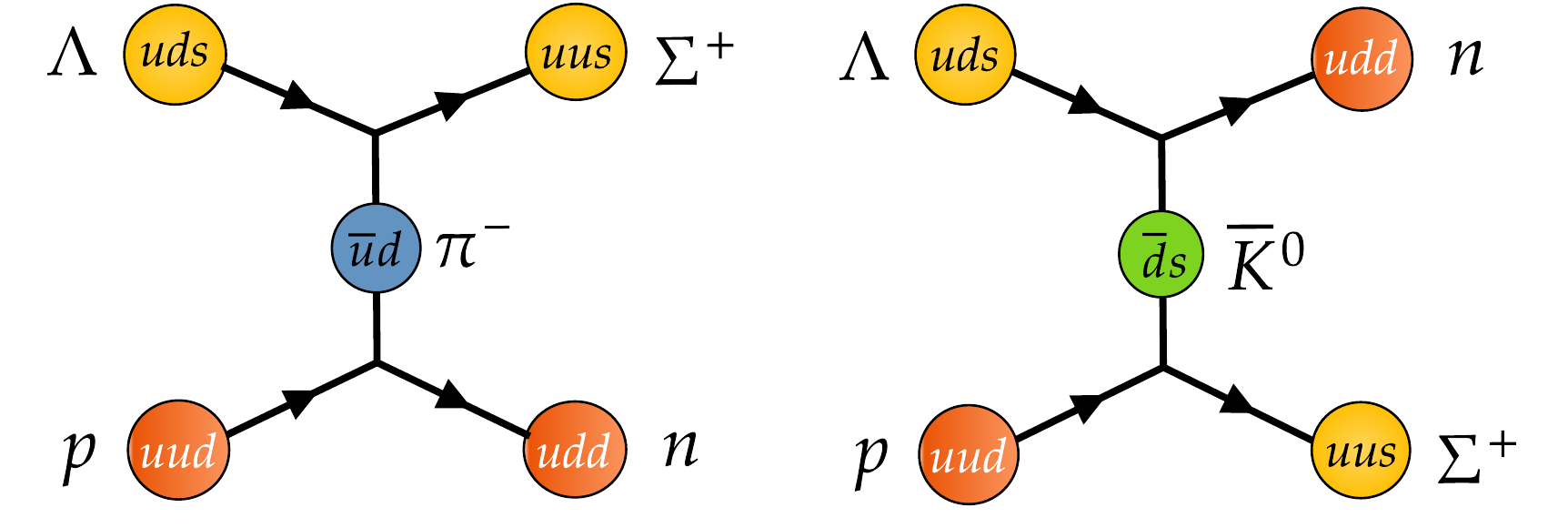} }
        \caption{(a) An overview and (b) Feynman diagram for the interaction between $\Lambda$ hyperon and Be nuclei. $\Sigma^+$ hyperon is produced in this process.}
        \label{fig:Feynman}
\end{figure}

As shown in Fig.~\ref{fig:r99bes}, hyperon pairs or final states including hyperons are produced by the collision of $e^+ e^-$ inside the beam pipe and will fly in the direction of the momentum. Some of the hyperons are able to arrive at the beam pipe or the inner wall of MDC, which are the target of this work, before decaying and scattering elastically or inelastically with the nucleus inside the material. For an example at BESIII, $\Lambda$ hyperons may interact with the Be nucleus inside the beam pipe and subsequently converse into $\Sigma^+$ by exchanging a $\pi$ or $K$ meson with the nuclei. At the same time, Be nuclei will converse into another nuclide. Fig.~\ref{fig:Feynman} is an overview and the Feynman diagrams for examples of such processes without covering all possible meson exchanges.

\begin{figure}[htbp]
	\centering
	\includegraphics[width=0.49\textwidth]{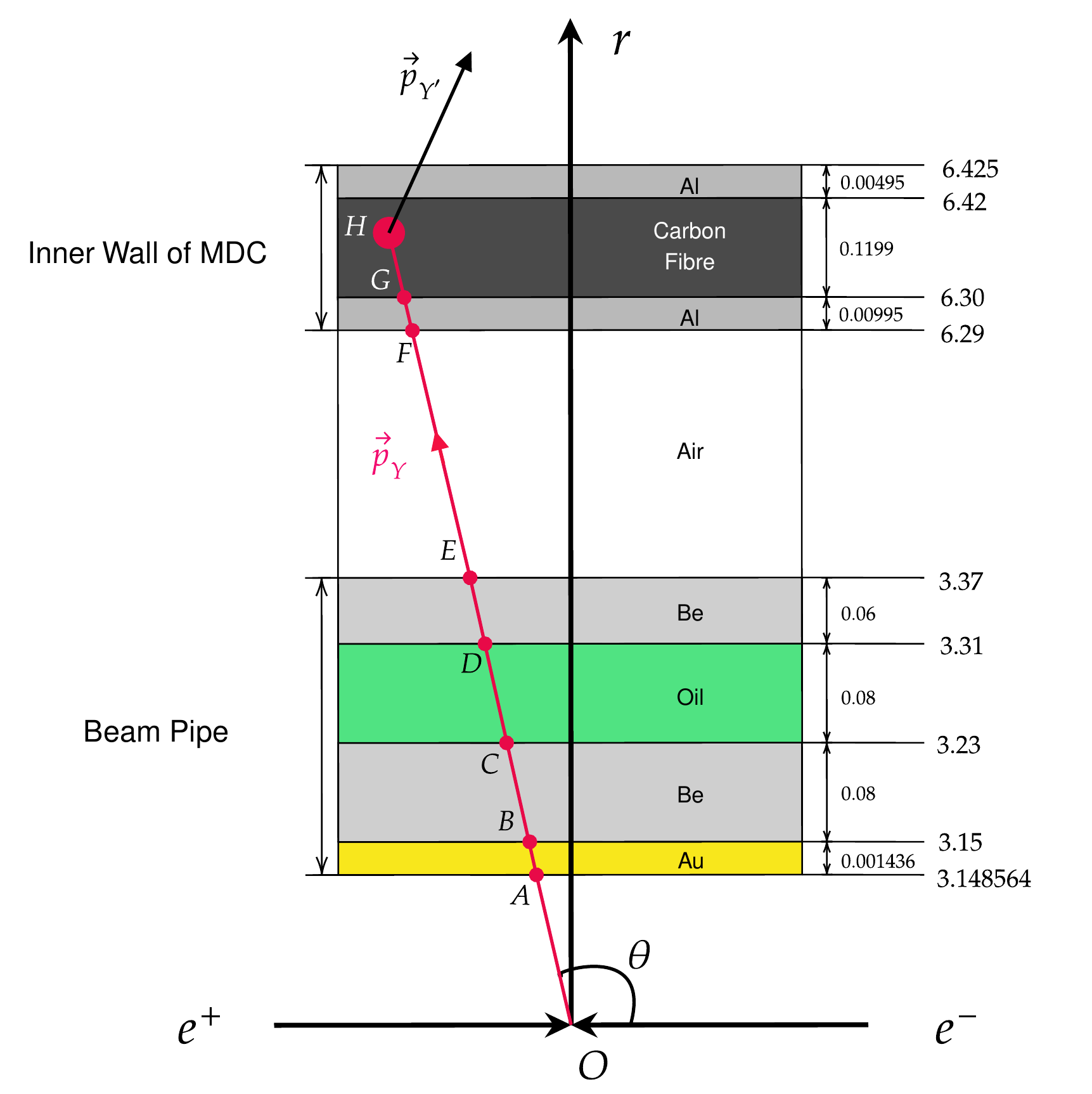}
	\caption{Target structure and the path length of $Y$ inside the target. The target consists of multiple layers composed of gold, beryllium, oil, aluminum and carbon fiber. $O$ is the interaction point of $e^+ e^-$ collision. The horizontal axis is the $e^+e^-$ beam line and the vertical axis (r-axis) denotes the distance away from the beam line. The position and thickness of each layer are listed in the figure, where the unit is centimeter. $\theta$ is the angle between incident $Y$ and z-axis.}
	\label{fig:interaction}
\end{figure}

The structure of the target used in this proposal is presented in Fig.~\ref{fig:interaction}. The target in this work consists of multiple layers composed of gold ($^{197}{\rm Au}$), beryllium ($^{9}{\rm Be}$), oil ($m_{^{12}{\rm C}} : m_{^{1}{\rm H}} = 84.923\% : 15.077\%$), aluminum ($^{27}{\rm Al}$) and carbon fiber ($m_{^{12}{\rm C}} : m_{^{1}{\rm H}} : m_{^{16}{\rm O}} = 69.7\% : 0.61\% : 29.69\%$), where $m_{\rm A}$ means the mass fraction of the certain nucleus. $O$ is the interaction point of $e^+ e^-$ collision. The vertical axis (z-axis) is the direction of the $e^+$ beam and the horizontal axis (r-axis) denotes the distance away from z-axis. $r_{i}$ and $t_{i}$ refer to the radius and thickness of each layer. Hyperon $Y$ is emitted in the direction of the polar angle $\theta$ and will react with the nucleus in the target materials at the scattering point $H$, resulting in the production of particle $Y^{\prime}$, so $AB$, $BC$, ..., $GH$ is the track length of $Y$ in each layer, of which the sum will be the total track length inside the target. And the corresponding density and molar mass are shown in Eq.~(\ref{equ:thickness}). Eq.~(\ref{equ:density}), where $\rho_{T}$ and $M$ refer to density and molar mass, respectively. The molar mass of oil and carbon fiber in the target is calculated by averaging the molar mass of C, H and O nucleus weighted by the number of each kind of nucleus inside the unit volume.

\begin{equation}
        \label{equ:thickness}
	\doublespacing
        r_{i}\left\{{
                \begin{array}{l}
                        r_1 = 3.148564~{\rm cm} \\
                        r_2 = 3.15~{\rm cm}     \\
                        r_3 = 3.23~{\rm cm}     \\
                        r_4 = 3.31~{\rm cm}     \\
                        r_5 = 3.37~{\rm cm}     \\
                        r_6 = 6.29~{\rm cm}     \\
                        r_7 = 6.30~{\rm cm}     \\
                        r_8 = 6.42~{\rm cm}     \\
                        r_9 = 6.425~{\rm cm}
                \end{array}}
                \right.
                t_{i}\left\{{
                        \begin{array}{l}
                                t_1 = 0.001436~{\rm cm} \\
                                t_2 = 0.08~{\rm cm}     \\
                                t_3 = 0.08~{\rm cm}     \\
                                t_4 = 0.06~{\rm cm}     \\
                                t_5 = 0.00995~{\rm cm}  \\
                                t_6 = 0.1199~{\rm cm}   \\
                                t_7 = 0.00495~{\rm cm}  \\
                        \end{array}}
                        \right.
\end{equation}

\begin{widetext}
        \begin{equation}
                \label{equ:density}
		\doublespacing
                \rho_T,~M\left\{{
                        \begin{array}{lll}
                                \rho({\rm Au}) = 19.32~{\rm g \cdot cm^{-3}},   &       M({\rm Au}) = 197~{\rm g \cdot mol^{-1}},       &       r_1 \le r \le r_2       \\
                                \rho({\rm Be}) = 1.848~{\rm g \cdot cm^{-3}},   &       M({\rm Be}) = 9~{\rm g \cdot mol^{-1}},       &       r_2 \le r \le r_3        \\
                                \rho({\rm Oil}) = 0.81~{\rm g \cdot cm^{-3}},   &       M({\rm Oil}) = 4.51~{\rm g \cdot mol^{-1}},       &       r_3 \le r \le r_4        \\
                                \rho({\rm Be}) = 1.848~{\rm g \cdot cm^{-3}},   &       M({\rm Be}) = 9~{\rm g \cdot mol^{-1}},       &       r_4 \le r \le r_5        \\
                                \rho({\rm Al}) = 2.7~{\rm g \cdot cm^{-3}},   &       M({\rm Al}) = 27~{\rm g \cdot mol^{-1}},       &       r_6 \le r \le r_7        \\
                                \rho({\rm Carb}) = 1.57~{\rm g \cdot cm^{-3}},   &       M({\rm Carb}) = 12.09~{\rm g \cdot mol^{-1}},       &       r_7 \le r \le r_8        \\
                                \rho({\rm Al}) = 2.7~{\rm g \cdot cm^{-3}},   &       M({\rm Al}) = 27~{\rm g \cdot mol^{-1}},       &       r_8 \le r \le r_9        \\
                        \end{array}}
                        \right.
        \end{equation}
\end{widetext}

\subsection{Monte Carlo simulation of the signal processes}
\label{sec:signal_mc}

To study for detail the potential measurements at BESIII, MC samples of the decays of $J/\psi(\psi(3686))$ to hyperon-antihyperon pairs $Y\bar{Y}$ ($Y$ and $\bar{Y}$ represent hyperon and antihyperon), including $\Sigma^{+}\bar{\Sigma}^{-}$, $\Sigma^{-}\bar{\Sigma}^{+}$, $\Sigma^{0}\bar{\Sigma}^{0}$, $\Xi^-\bar{\Xi}^{+}$, $\Xi^{0}\bar{\Xi}^{0}$ and $\Omega^- \bar{\Omega}^{+}$, are generated.

The angular distribution can be described as~\cite{Faldt:2017kgy, Perotti:2018wxm}
\begin{equation}
        \label{equ:angular_distri}
        \frac{{\rm d}N(Y)}{{\rm dcos}\theta} \propto 1+\alpha_{\psi}{\rm cos}^{2}\theta,
\end{equation}
where $\alpha_{\psi}$ is the parameter of the angular distribution measured by experiment~\cite{BESIII:2022qax, BESIII:2021ypr, BESIII:2020fqg, BESIII:2018cnd, BESIII:2022lsz, BESIII:2023drj} and $\theta$ denotes the angle between $Y$ hyperon and the direction of $e^+$ beam. Considering hyperons will decay along the track and may not be able to arrive at the beam pipe or the inner wall of MDC, the probability for hyperons from $x_{0}$ to be alive at $x$ is given by
\begin{equation}
        \label{equ:decay_length}
        P(x) = {\rm exp}[-\frac{M}{p}\frac{x-x_0}{\tau}],
\end{equation}
where $M$, $p$ and $\tau$ are the rest mass, momentum and intrinsic lifetime of hyperon $Y$, respectively.

\begin{figure}[htbp]
	\centering
	\includegraphics[width=0.45\textwidth]{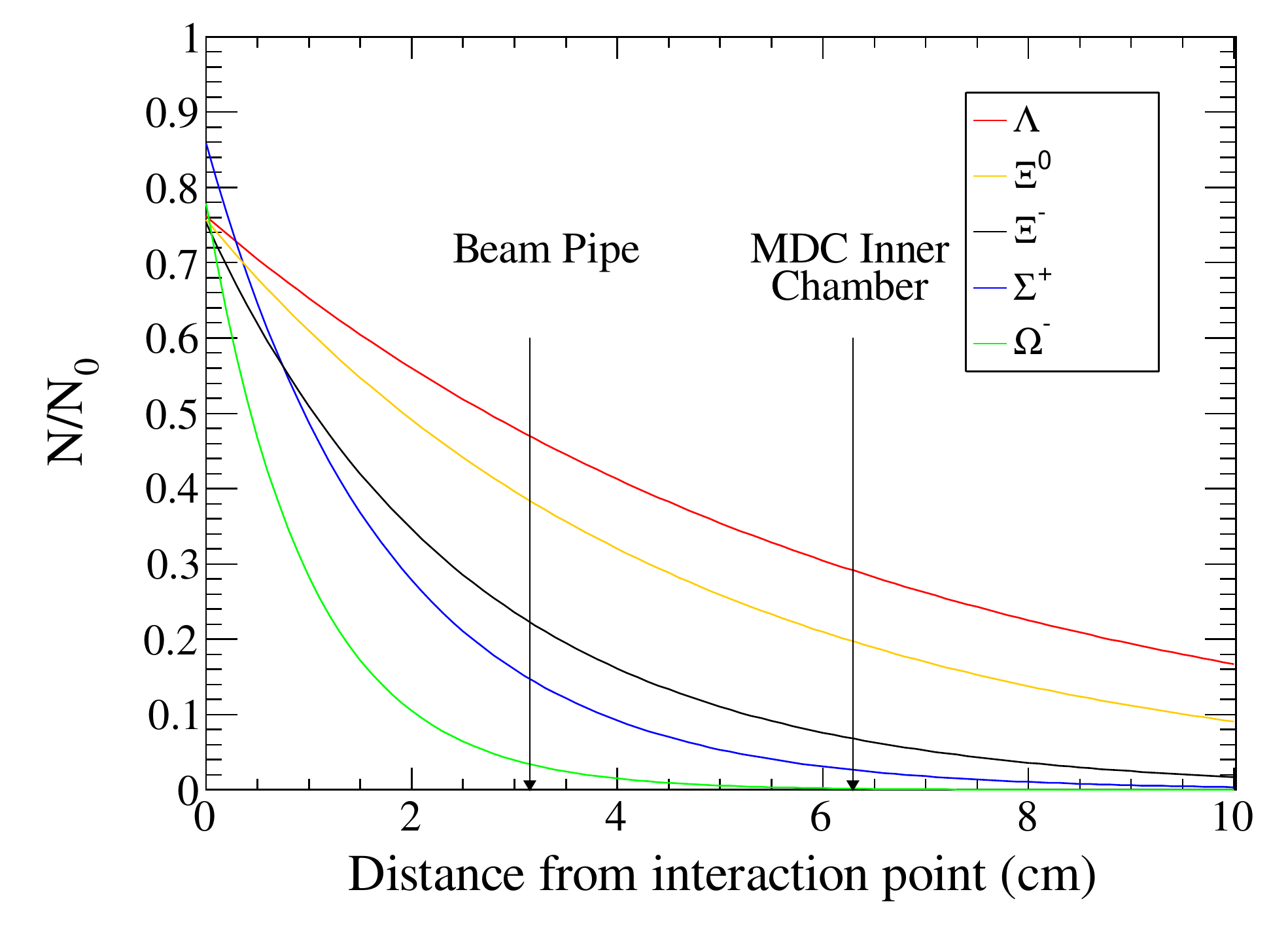}
	\caption{The ratio of surviving hyperons over the total number of hyperons produced by charmonium decay. Only hyperons with $|{\rm cos}\theta|<0.8$ are recorded considering the acceptance range and detection efficiency of BESIII spectrometer.}
	\label{fig:decay_ratio}
\end{figure}

After considering the acceptance range of the BESIII detectors, we get the ratio of surviving hyperons over the total number of hyperons produced by charmonium decay as shown in Fig.~\ref{fig:decay_ratio}. We notice that the start points of the curves of different hyperons are not the same, because different hyperons have different angular distributions when they are produced and we only record hyperons with $|{\rm cos}\theta|<0.8$ considering the acceptance of the BESIII spectrometer.

The reactions between the hyperon or antihyperon and the material of BESIII spectrometer are simulated by the the official physics list $\rm QGSP\_BERT$ defined by Geant4~\cite{Geant4:PhysicsList}, in which the hyperon is simulated by the Bertini intra-nuclear cascade model~\cite{Apostolakis:2009egq, Heikkinen:2003sc}. The implementation of this model into Geant4 employs many of the standard intra-nuclear cascade features developed by Bertini and Guthrie~\cite{Bertini:1971xb}:
\begin{itemize}
        \item[*] classical scattering without matrix elements
        \item[*] free hadron-nucleon cross sections and angular distributions which are taken from experiment
        \item[*] step-like nuclear density distributions and potentials
\end{itemize}
The second feature, in principle, allows the model to be extended to any particle for which there are sufficient double-differential cross section measurements, but also brings difficulties in the simulation of processes that are rarely measured by experiments like the inelastic scattering of hyperon and nucleon.

In the simulation, the decay of the incident hyperon has been considered using the measured lifetime, so the hyperons may decay before entering the target or inside the target if they don't interact with the target, which will finally be reflected in the calculation of $\mathcal{L}_{Y}$ as described in Sec.~\ref{sec:method}.

\subsection{The method to study the hyperon-nuclei interactions at BESIII}
\label{sec:method}

As shown in Fig.\ref{fig:r99bes}, considering the process $\psi \to XY,~YA \to Y^{\prime} A^{\prime}$ ($Y$ is the hyperon that interacts with nucleons, $X$ is the other particles except $Y$ produced in the decay of charmonia, $Y^{\prime}$ is the new hyperon created by the hyperon-nucleon interaction, $A$ and $A^{\prime}$ are the nucleus inside the target before and after the interaction), an experimental method called ``Double Tag" method can be used in such measurements. 

The accurate information of the hyperon $Y$ that interacts with material, including momentum and direction, can be obtained by reconstructing the other particles denoted as $X$ among the final state, which is called ``single-tag". The yield of the hyperons of interest $N_{\rm ST}$ is obtained by fitting the recoil mass ($RM_{X}$) distribution of $X$, which is defined as
\begin{equation}
	RM_{X} = \sqrt{|p_{e^+ e^-} - p_{X}|^{2}},
\end{equation}
where $p_{e^+ e^-}$ and $p_{X}$ are the 4-momenta of the $e^+ e^-$ and $X$. $RM_{X}$ distribution will have a peak around the intrinsic mass of $Y$.

Then the hyperon produced by the hyperon-nucleon scattering denoted as $Y^{\prime}$ are reconstructed by the final states of its decay, which is called ``double-tag". The yield of $Y^{\prime}$ denoted as $N_{\rm DT}$ is obtained by fitting the distribution of the invariant mass of $Y^{\prime}$, which can be expressed as
\begin{equation}
	\label{equ:n_dt}
	N_{\rm DT}={\mathcal L}_{Y} \cdot \sigma(YA \rightarrow Y^{\prime}A^{\prime}) \cdot {\mathcal B}(Y^{\prime}) \cdot \epsilon_{\rm sig},
\end{equation}
where $\sigma(YA \rightarrow Y^{\prime}A^{\prime})$ is the cross section of the hyperon-nucleon process of interest, ${\mathcal B}(Y^{\prime})$ is the branching fraction of the decay channel used to reconstruct $Y^{\prime}$ and $\epsilon_{\rm sig}$ is the selection efficiency of $Y^{\prime}$ for the specific decay channel obtained from signal MC sample. To finally determine the cross section of such processes, a specially defined variable $\mathcal{L}_{Y}$, named ``effective luminosity'', is introduced to account for the properties of the target and the behavior of the incident hyperon beam and estimated using signal Monte Carlo (MC) samples described in Sec.~\ref{sec:signal_mc}.

The formula used to calculate $\mathcal{L}_{Y}$ is given by
\begin{equation}
        \label{equ:effective_lum}
        \mathcal{L}_{Y} = N_{\rm ST} \cdot \frac{N_{A} \cdot \rho_{T} \cdot l}{M},
\end{equation}
where $N_{\rm ST}$ is the number of single-tagged events, $N_{A}$ is the Avogadro constant, $\rho_{T}$ is the density of the target, $l$ is the average path length of the $Y$ beam inside the target and $M$ is the molar mass of the target. In the proposal of this work, the target is composed of several layers of different materials. The total value of $\mathcal{L}_{Y}$ is the sum of the contributions of each layer, which will be
\begin{equation}
        \label{equ:total_lum}
        \mathcal{L}_{Y} = \sum_{j}^{7} \mathcal{L}_{Y}^{j} = N_{\rm ST} \cdot N_{A} \cdot \sum_{j}^{7} \frac{\rho_{T}^{j} \cdot l^{j}}{M^{j}} \cdot \mathcal{R}_{\sigma}^{j},
\end{equation}
where $j$ is the index of the layers. Since the contribution from each layer and the cross section for different kinds of nuclei are not the same, a ratio of the scattering cross section (${\mathcal R}_{\sigma}$) of incident $Y$ and each kind of material is necessary. According to the previous studies, scattering happens mostly with single nucleons on the nucleus surface in the case of low and intermediate energy~\cite{Barton:1982dg, Cooper:1995ix, WA89:1997dmp, Botta:2001fu, Astrua:2002zg, Lee:2018epd}. If we define the effective proton number ($Z_{\rm eff}$) and effective neutron number ($N_{\rm eff}$) as the number of nucleons inside a single nuclei that are possible to be interact with the incident hyperon, $\mathcal{R}_{\sigma}$ will be proportional to $Z_{\rm eff}$ or $N_{\rm eff}$ under the scenario above, calculated as $\mathcal{R}_{\sigma} \propto Z_{\rm eff} = A^{\frac{2}{3}} \times \frac{Z}{A} = \frac{Z}{A^{\frac{1}{3}}}$ for the interactions with protons and $\mathcal{R}_{\sigma} \propto N_{\rm eff} = A^{\frac{2}{3}} \times \frac{N}{A} = \frac{N}{A^{\frac{1}{3}}}$ for the interactions with neutrons, where $A$, $Z$ and $N$ are the number of nucleons, protons and neutrons in a single nuclei. Taking $J/\psi \to \Lambda \bar{\Lambda}, \bar{\Lambda} \to \bar{p} \pi^+, \Lambda A \to \Sigma^+ X$ as an example, $\mathcal{R}_{\sigma}$ for each layer will be
\begin{equation}
        \label{equ:assum_2}
	\doublespacing
        {\mathcal R}_{\sigma}^{j}\left\{{
                \begin{array}{l}
                        {\mathcal R}_{\sigma}^{1} = 7.06,         \\
                        {\mathcal R}_{\sigma}^{2} = 1.0,          \\
                        {\mathcal R}_{\sigma}^{3} = 0.789,        \\
                        {\mathcal R}_{\sigma}^{4} = 1.0,          \\
                        {\mathcal R}_{\sigma}^{5} = 2.253,        \\
                        {\mathcal R}_{\sigma}^{6} = 1.365,        \\
                        {\mathcal R}_{\sigma}^{7} = 2.253,        \\
                \end{array}}
                \right.
\end{equation}
after normalizing the cross sections for each layer to the largest component of the target $^{9}{\rm Be}$. The ratio for oil and carbon fiber is the average of the ratios of C, H and O weighted by the total number of the corresponding nucleus inside the unit volume, which is described as
\begin{equation}
        \label{equ:ratio_mix}
        \mathcal{R}_{\sigma}^{\rm mixture} = \frac{\mathcal{R}_{\sigma}^{C}N^{C} + \mathcal{R}_{\sigma}^{H}N^{H} + \mathcal{R}_{\sigma}^{O}N^{O}}{N^{C} + N^{H} + N^{O}},
\end{equation}
where $N^{C}$, $N^{H}$ and $N^{O}$ are the numbers of C, H and O nucleus inside the unit volume, respectively.

The average path length $l^{j}$ inside each layer is calculated using signal MC sample of $YA \rightarrow Y^{\prime}A^{\prime}$, in which the decay of the $Y$ beam is taken into consideration as described in detail in Sec.~\ref{sec:signal_mc}. $l^{j}$ is calculated event by event as
\begin{equation}
        \label{equ:average_path_length}
        l^{j} = \frac{\sum_{i}^{N_{\rm ST}^{\rm MC}} l^{ij}}{N_{\rm ST}^{\rm MC}},
\end{equation}
where $N_{\rm ST}^{\rm MC}$ is the total number of the single-tagged events in the MC sample and $l^{ij}$ refers to the path length of the $Y$ of the $i_{\rm th}$ event inside the $j_{\rm th}$ layer. As shown in Fig.~\ref{fig:interaction}, $H$ is the scattering points of $Y$ and nucleon so that $AB$, $BC$, ..., $GH$ are the track length of $Y$ in each layer that is obtained from the true information of signal MC sample and will be 0 if $Y$ hyperon doesn't enter the layer. The sum of all the segments is the total track length inside the target. For the case of $J/\psi \to \Lambda \bar{\Lambda}, \bar{\Lambda} \to \bar{p} \pi^+, \Lambda {^{9}{\rm Be}} \to \Sigma^+ {^{9}{\rm Li}}$, $l^{j}$ will be
\begin{equation}
	\doublespacing
        \label{equ:average_length}
        l^{j}\left\{{
                \begin{array}{l}
                        l^{1} = 0.10 \times 10^{-3} ~{\rm cm},         \\
                        l^{2} = 5.76 \times 10^{-3} ~{\rm cm},          \\
                        l^{3} = 5.68 \times 10^{-3} ~{\rm cm},        \\
                        l^{4} = 4.21 \times 10^{-3} ~{\rm cm},          \\
                        l^{5} = 0.43 \times 10^{-3} ~{\rm cm},        \\
                        l^{6} = 5.13 \times 10^{-3} ~{\rm cm},        \\
                        l^{7} = 0.21 \times 10^{-3} ~{\rm cm},        \\
                \end{array}}
                \right.
\end{equation}
In the calculation of the average path length, the decay of incident hyperons and the potential interactions with the target material have both been taken into account. Since the rate of hyperon-nucleus interactions is about 1\textperthousand of the hyperon decay inside the target, $l_{j}$ will be influenced very slightly by the hyperon-nucleus interactions, however, compared to the final statistical uncertainty, this impact is negligible.

Combining Eq.~(\ref{equ:total_lum}) and Eq.~(\ref{equ:average_path_length}), $\mathcal{L}_{Y}$ will be
\begin{equation}
        \label{equ:final_lum}
        \mathcal{L}_{Y} = N_{\rm ST} \cdot \frac{N_{A}}{N_{\rm ST}^{\rm MC}} \cdot \sum_{j}^{7} \sum_{i}^{N_{\rm ST}^{\rm MC}} \frac{\rho_{T}^{j} \cdot l^{ij}}{M^{j}} \cdot \mathcal{R}_{\sigma}^{j}.
\end{equation}

$\mathcal{L}_{Y}$ contains the information of both the number of incident hyperons and the properties of the target. The beam intensity, or the single-tagged events $N_{\rm ST}$, cannot be obtained precisely in this paper while it will be obtained from the fit in the data analysis of BESIII. In this paper, we only provide the value of $\mathcal{L}_{Y}/N_{\rm ST}$ that can be calculated using the MC samples and is independent from the data of the BESIII experiment.

Using the signal MC sample, we calculate the $\mathcal{L}^{j}_{\Lambda}/N_{\rm ST}$ for $\Lambda {^{9}{\rm Be}} \to \Sigma^+ {^{9}{\rm Li}}$ reaction of each layer as is shown in Eq.~(\ref{equ:lum_layer}). For the whole target, $\mathcal{L}_{\Lambda}/N_{\rm ST} = 23.59 \times 10^{21}~{\rm cm^{-2}}$. $\mathcal{L}_{Y}/N_{\rm ST}$ of other hyperons can be found in Tab.~\ref{tab:hyper}.
\begin{equation}
	\doublespacing
        \label{equ:lum_layer}
	{\mathcal L}_{\Lambda}^{j}/N_{\rm ST}\left\{{
                \begin{array}{l}
			\mathcal{L}_{\Lambda}^{1}/N_{\rm ST} = 0.43 \times 10^{21} ~{\rm cm^{-2}},         \\
                        \mathcal{L}_{\Lambda}^{2}/N_{\rm ST} = 7.12 \times 10^{21} ~{\rm cm^{-2}},          \\
                        \mathcal{L}_{\Lambda}^{3}/N_{\rm ST} = 4.85 \times 10^{21} ~{\rm cm^{-2}},        \\
                        \mathcal{L}_{\Lambda}^{4}/N_{\rm ST} = 5.21 \times 10^{21} ~{\rm cm^{-2}},          \\
                        \mathcal{L}_{\Lambda}^{5}/N_{\rm ST} = 0.58 \times 10^{21} ~{\rm cm^{-2}},        \\
                        \mathcal{L}_{\Lambda}^{6}/N_{\rm ST} = 5.12 \times 10^{21} ~{\rm cm^{-2}},        \\
                        \mathcal{L}_{\Lambda}^{7}/N_{\rm ST} = 0.28 \times 10^{21} ~{\rm cm^{-2}},        \\
                \end{array}}
                \right.
\end{equation}

Combining Eq.~(\ref{equ:n_dt}), the $\sigma(YA \rightarrow Y^{\prime}A^{\prime})$ is obtained to be
\begin{equation}
        \label{equ:cross_section}
	\sigma(YA \rightarrow Y^{\prime}A^{\prime}) = \frac{N_{\rm DT}}{\epsilon_{\rm sig} \cdot {\mathcal{L}_{Y}}} \cdot \frac{1}{{\mathcal B}(Y^{\prime})}.
\end{equation}

\subsection{Expected signal yield at BESIII}

Up to now, 10 billion $J/\psi$ and 2.7 billion $\psi(3686)$ events have been collected at BESIII, in which tens of millions of ground-state hyperons are produced. Not only could rare decays of hyperons be precisely measured~\cite{Li:2016tlt}, but also the interactions between hyperons and nucleons can be studied.

Using the MC samples described in Sec.~\ref{sec:signal_mc} and the method in Sec.~\ref{sec:method}, we estimate the potential measurement of the hyperon-nucleus interactions at BESIII in Tab.~\ref{tab:hyper} with assuming the cross section of such interactions to be $20~{\rm mb}$. Only hyperon pairs from charmonium decay are calculated, but our method can also be applied to other hyperon-involved multi-body final states. There will be as many as several thousand scattering events at BESIII before reconstruction. Taking the reconstruction efficiency of different hyperon-antihyperon pairs to be $5\% \sim 30\%$, there will be a considerable quantity of signals in the data analysis.

\begin{table*}[htbp]
	\caption{Expected signal yields of hyperon-nucleon scattering at BESIII considering the acceptance range and selection efficiency of BESIII. $p_{\rm max}$ is the maximum momentum of the antihyperon; $\mathcal{B}_{decay}$ is the branching fraction of the given decay channel of charmonium; $n^Y_{\rm BP}$ is the number of the tagged antihyperons reaching the beam pipe; $\mathcal{B}_{tag}$ is the branching fraction of the decay channel used in the single-tag side; and $\mathcal{L}_{Y}$ is the effective luminosity of the hyperon beam. The last column shows the expected signal yields for different hyperons before reconstruction. Predictions for STCF are also listed in this table.}
    \label{tab:hyper}
    \centering
	\renewcommand{\arraystretch}{1.2}
	\begin{tabular}{p{1.5cm}<{\centering}|p{1cm}<{\centering}|p{3.0cm}<{\centering}|p{2.0cm}<{\centering}|p{1.5cm}<{\centering}|p{2.0cm}<{\centering}|p{1.5cm}<{\centering}|p{2.3cm}<{\centering}|p{2.0cm}<{\centering}}
	    \hline
	    \hline
	    Hyperon    &$c\tau$ (cm)  &decay mode                               &${\cal B}_{decay}$~\cite{ParticleDataGroup:2022pth} ($\times 10^{-3}$) &$p_{\rm max}$ (MeV/$c$)  &$n^Y_{\rm BP}$ ($\times 10^5$ for BESIII or $\times 10^8$ for STCF) & $\mathcal{B}_{tag}~(\%)$	&	$\mathcal{L}_{Y}/N_{\rm ST}~(10^{21}\cdot {\rm cm^{-2}})$	&	Estimated signal yield ($\times 10^3$ for STCF) \\ \hline
		$\Lambda$  &7.89          &$\jpsi\to \Lambda\bar{\Lambda}$          &$1.89\pm0.09$                         &1074                     &26 	&64	&23.59		&5290	\\

		$\Sigma^+$ & 2.40 & $\jpsi\to \Sigma^+ \bar{\Sigma}^-$    &  $1.07\pm0.04$ &  992  & 4 &52	&4.83	&537	\\

		$\Xi^0$ & 8.71 & $\jpsi\to \Xi^0 \bar{\Xi}^0$ &  $1.17\pm0.04$   &  818 & 7 &64	&	15.81	&2368	\\ 

		$\Xi^-$ & 4.91 & $\jpsi\to \Xi^- \bar{\Xi}^+$ &  $0.97\pm0.08$   &  807 & 3 &64	&7.44	&924	\\ 

		$\Omega^-$ & 2.46 & $\psip\to \Omega^-\bar{\Omega}^+$ &  $0.056\pm0.003$   &  774 & 0.05 &43	&2.61	&3 \\ 
      \hline
	    \hline
	\end{tabular}
\end{table*}

\section{Towards the interactions between hyperon and single nucleon}
\label{sec:effective_lum}


It should be pointed out that the raw measurements at BESIII are the hyperon-nuclei instead of the hyperon-nucleon cross sections, of which the latter is more important for the theoretical studies. The corresponding hyperon-nucleon cross sections can be extracted by calculating the effective nucleon number. In Section~\ref{sec:method}, a scenario has been provided to calculate the effective nucleon number with assuming the cross section is proportional to the number of nucleons in the nucleus surface in the case of low and intermediate energy~\cite{Barton:1982dg, Cooper:1995ix, WA89:1997dmp, Botta:2001fu, Astrua:2002zg, Lee:2018epd}. Using the ratio of the effective nucleon number of multiple materials that is denoted as $\mathcal{R}_{\sigma}$ in Section~\ref{sec:method} and the provided $\mathcal{L}_{Y}/N_{\rm ST}$ for each layer, the measured cross section can be normalized to any kind of nucleus, naturally including that with a single nucleon.

Here we provide another scenario to calculate the effective nucleon number based on a assumption with quasifree scattering process. We consider the Eikonal Approximation~\cite{Bando:1990yi,Joachain:1975,Cooper:1995ix}, with which the connection between hyperon-nuclei and hyperon-nucleon cross sections can be established via the effective nucleon number calculated by

	\begin{widetext}
\begin{eqnarray}
& N_{\mathrm{eff}}\left(Z_{\mathrm{eff}}\right) = \frac{N(Z)}{A}\int \rho(\mathbf{r}) \exp \left\{-\bar{\sigma}_{i} \int_{-\infty}^z \rho\left(x, y, z^{\prime}\right) d z^{\prime}-\right. \left.\bar{\sigma}_{f} \int_z^{\infty} \rho\left(x, y, z^{\prime}\right) d z^{\prime}\right\} d^3 \mathbf{r}
\end{eqnarray}
	\end{widetext}



Here $N$ ($N_{\rm eff}$) and $Z$ ($Z_{\rm eff}$) are the (effective) neutron number and proton number of the nuclei. $A$ is the atom mass. $\rho$ is the nuclear density distribution with $A = \int \rho(\mathbf{r}) d\mathbf{r}$. $\bar{\sigma}_i$ and $\bar{\sigma}_f$ are the isospin-averaged cross sections of the elastic scatterings between the two particles of the incoming and outgoing states, respectively. When the momentum of the incident hyperon is larger, the hyperon-nucleon scattering cross section has an approximate relationship with the hyperon-nuclei cross section as: $\sigma_{Y-A} = \sigma_{Y-n} \times N_{\mathrm{eff}}(Z_{\mathrm{eff}})$.

	As an example, we here consider to calculate the cross section of $\Lambda + p\to \Sigma^+ + n$ from the recently measured cross section of $\Lambda + {^{9}{\rm Be}} \to \Sigma^+ + X$ by BESIII to be $\sigma = (37.3 \pm 4.7 \pm 3.5)~{\rm mb}$~\cite{BESIII:2023trh}. The cross section for the corresponding hyperon-nucleon scattering $\Lambda + p \to \Sigma^+ + n$ is calculated using the scenario in Section~\ref{sec:method} to be $\sigma(\Lambda + p \to \Sigma^+ + n) = (19.3 \pm 2.4 \pm 1.8)~{\rm mb}$.

	The nuclear density of ${^9{\rm Be}}$ is depicted by the two-parameter parameterization of the Fermi distribution as~\cite{DeJager:1974liz}:
\begin{equation}\label{rho-nuclear}
\rho(r) = \rho_0\left[1+\alpha\left(\frac{r}{a}\right)^2\right] \operatorname{Exp}\left[-\left(\frac{r}{a}\right)^2\right],
\end{equation}
where $\alpha = 0.63$~fm and $a = 1.77$~fm. $\rho_{0}$ is the normalization factor.

To calculate the effective proton number in the ${^{9}{\rm Be}}$ nucleus, there are two scenarios using either the experimentally measured or the theoretically predicted elastic cross sections of $\Lambda p$ and $\Sigma^+ n$.

Firstly, based on the measurements in Ref.~\cite{KEK-PSE289:2005nsj,CLAS:2021gur}, we can roughly obtain the mean elastic cross sections to be $\bar{\sigma}(\Lambda p) \approx (17.8\pm2.4)$ $\mathrm{mb}$ and $\bar{\sigma}(\Sigma^+ n) \approx 51.3_{-25.5}^{+51.7}~\mathrm{mb}$ with the momentum of $\Lambda/\Sigma$ Hyperon around 1~GeV/c$^2$. Hence, we can evaluate the effective proton number to be $Z_{\mathrm{eff}} \in [1.1,2.1]$ for $\Lambda + {^{9}{\rm Be}} \to \Sigma^+ + {^{9}{\rm Li}}$ reaction, where the large uncertainty arises mostly from the measurements of $\Sigma$-nucleon elastic cross section. If we take the central value 1.6 as a reference, the cross section of $\Lambda p \to \Sigma^+ n$ can be calculated to be $\sigma(\Lambda p \to \Sigma^+ n) = (23.3 \pm 3.7 ^{+11.0}_{-5.5})~{\rm mb}$, where the first uncertainty is from the measured cross section and the second is induced by the uncertainty of $Z_{\mathrm{eff}}$.

The uncertainties will be significantly reduced if taking the theoretical calculated $\bar{\sigma}(\Lambda p) = (17.0 \pm 2.5)~{\rm mb}$ and $\bar{\sigma}(\Sigma^+ n) = (28.5\pm 2.5)~{\rm mb}$ from $\chi$EFT methods~\cite{Haidenbauer:2023qhf, Li:2016mln}. Using the above method, one can estimate the effective proton number to be $Z_{\mathrm{eff}} \in [1.9,2.2]$. Taking the central value 2.1 as a reference, the cross section of $\Lambda p \to \Sigma^+ n$ can be calculated to be $\sigma(\Lambda p \to \Sigma^+ n) = (17.8 \pm 2.8 ^{+1.8}_{-0.8})~{\rm mb}$, where the first uncertainty is from the measured cross section and the second is induced by the uncertainty of $Z_{\mathrm{eff}}$.

Using the input elastic cross sections from no matter experimental measurements or theoretical calculation, the cross section of $\Lambda + p \to \Sigma^+ + n$ reaction calculated under the scenario described above is consistent with the cross section calculated under the scenario in Section~\ref{sec:method}. In future, the uncertainty of the effective nucleon number will be reduced by the proposed measurements of elastic cross sections by BESIII.

\section{Discussions} 
\label{sec:disc}

Differential cross sections of hyperon-nucleus interactions can also be extracted in our proposed method. For the process $\psi \to XY,~YA \to Y^{\prime} A^{\prime}$, the momentum of $Y$ can be obtained by studying the recoiling system of the tagged hyperon $X$ so that the angle between $Y^{\prime}$ and $Y$ will be available. With the momentum of $Y^{\prime}$, the differential cross sections can be measured within the whole phase space if the statistics are large enough.

Furthermore, the polarization of the hyperon pairs from the decay of charmonia produced at BESIII has been studied thoroughly from both experimental~\cite{BESIII:2022qax, BESIII:2021ypr, BESIII:2020fqg, BESIII:2018cnd, BESIII:2022lsz, BESIII:2023drj} and theoretical~\cite{Faldt:2017kgy} sides in recent years. It is found that the polarization will be a function of the polar angle of the hyperons, which inspires us to suppose that the relationship between the cross sections and the polarization of the incident hyperons can also be studied in detail by constraining the phase space. The polarization-dependent mechanism for such processes will be of great significance for figuring out the role of spin in hyperon-nucleon interactions and the potential of strong interaction. Meanwhile, large polarization effects have been observed in the produced hadrons in the study of elastic hadronic scattering. Similarly, the polarization of the produced hyperons in hyperon-nucleon interactions will also be a possible topic for the research at BESIII by measuring the angular distribution of the particles from the hyperon decay, which will be potential clues for the spin-involved interaction between hyperon and nucleon.

Similar to hyperons, the interactions between antihyperons and nucleus/nucleons, including scattering and annihilation, can also be studied by tagging the hyperons produced in pair. This will provide very important information for the theoretical research given that the present measurements of such processes are far from enough.

In the near future, new generation of $e^+ e^-$ collider in proposal with a super high intensity beam called Super Tau-Charm Facility (STCF) will reach a peak luminosity of $1\times10^{35}~{\rm cm^{-2}s^{-1}}$~\cite{Achasov:2023gey}, which will be 100 times of the present BEPCII. Especially, the uncertainty of center-of-mass energy will improve from $1.2~{\rm MeV}$ to $20 \sim 80~{\rm keV}$ if monochromatic collision can be realized at the narrow resonances such as $J/\psi$~\cite{Telnov:2020rxp}, so that the cross section of $J/\psi$ production will further increase by 10 times. Given that there will be $10^{12} \sim 10^{13}$ $J/\psi$ produced at STCF per year, $10^6 \sim 10^7$ scattering events will be available according to the calculation in Tab.~\ref{tab:hyper}, which will significantly proceed the research on the hyperon-nuclei/nucleon interactions and help a lot in revealing the puzzle of the internal structure of neutron stars.

\section{Summary} 
\label{sec:summary}

This work provides a method to study hyperon-nuclei interactions by measuring the scattering between hyperons from charmonium decay at BESIII and the nucleus inside the dense matter in the beam pipe or other supporting structures and, subsequently, extracting the cross sections of corresponding hyperon-nucleon interactions via further phenomenological calculation. This method has been verified by the recent measurements from BESIII~\cite{BESIII:2023clq, BESIII:2023trh} and greatly broadens the frontier of physics at $e^+ e^-$ colliders. Thanks to the double-tag method, differential cross sections of such interactions can also be measured, which will provide exciting hints for the research on the potential of strong interaction and the origin of color confinement. Similarly, kaon-nuclei/nucleon or antihyperon-nuclei/nucleon interactions can also be studied by the same method utilizing the abundant kaon and antihyperon beams from the decay of quantities of charmonia. Specially for STCF in future, there will be millions of events of hyperon-nuclei scattering and much more events of kaon-nuclei scattering, which will definitely benefit a lot the precise probe of the hyperon-nucleon interactions and provide the essential constraints for fields including the internal structure of neutron stars and the unified model for baryon-baryon interactions.

\acknowledgments

The authors thank Prof. Ziyan Deng and Prof. Li-Sheng Geng for the fruitful discussions. This work is supported by the National Natural Science Foundation of China (NSFC) under Contracts Nos. 11935018, 11875054; Chinese Academy of Sciences (CAS) Key Research Program of Frontier Sciences under Contracts No. QYZDJ-SSW-SLH003.

\bibliography{biblio.bib}

\end{document}